\documentstyle[12pt,amsmath]{article}
\begin{document}

\title{Fractional Quantum Hall Effect 
and vortex lattices.}

\author{S. V. Iordanski}

\date{Landau Institute for Theoretical Physics, 
Russian Academy of Sciences, Moscow, 117334 Russia}

\maketitle

\begin{abstract}

It is demonstrated that all observed fractions at moderate Landau level fillings in
the quantum Hall effect can be obtained without recourse to the
phenomenological concept of composite fermions. The possibility to have
the special topologically nontrivial many-electron wave functions is
considered. Their group classification indicate the special values of
of electron density in the ground states separated by
a gap from excited energies.

\end{abstract}

73.43.-f

\vspace{0.5cm}

The experimental discovery of Integer Quantum Hall Effect (IQHE) by
K.Klitzing (1980)and Fractional Quantum Hall Effect (FQHE) by Tsui,Stormer
and Gossard (1982) was one of the most outstanding achievements in condensed
matter physics of the last century.Despite the fact that more than twenty 
years have elapsed since the
experimental discovery of quantum Hall Effect (QHE), the theory of
this phenomenon is far from being complete (see reviews [1, 2]).
This is primarily true for the Fractional Quantum Hall Effect (FQHE),
which necessitates the electron--electron interaction and can by no
means be explained by the one-particle theory, in contrast to the
IQHE. The most successful variational many-electron wave
function for explaining the 1/3 and other odd inverse fillings was constructed 
by Laughlin[3, 4]. The explanation of other observed fractions was obtained by
various phenomenological hierarchial schemes with construction of the "daughter"
states from the basic ones (Haldane 1983,Laughlin 1984, B.Halperin 1984).

 In those works, the approximation of extremely high magnetic
field was used, in which one is restricted to the states of
the lowest Landau level. However, this does not conform to the
experimental situation, where the cyclotron energy is on the order of
the mean energy of electron--electron interaction. Moreover, this
approach encounters difficulties in generalizing to the other
fractions. Computer simulations give a rather crude approximation 
for the realistic multiparticle functions, because the
number of particles in the corresponding calculations on modern
computers does not exceed several tens.

The most successful phenomenological description is given by the
Jain's model of "composite" fermions [5, 6], which predicts the
majority of observed fractions. According to this model, electrons
are dressed by magnetic-flux quanta with magnetic field 
concentrated in an infinitely narrow region around each electron. It
is assumed that even number of flux quanta provides that these
particles are fermions. The inclusion of this additional
magnetic field in the formalized theory leads to the so-called Chern--
Simons Hamiltonian. I describe this approach following close to [7].

One can perform canonical transfomation of the basis of the many particle wave
functions $\Phi\rightarrow\exp{i\hat{S}}\Phi$ where $\hat{S}$ a hermitian
operator:
\begin{equation}
\label{s}
\hat{S}=\int h({\bf \xi},{\bf \xi}')\psi^{+}({\bf \xi})\psi({\bf \xi})
\psi^{+}({\bf \xi}')\psi({\bf \xi}') d^2\xi d^2\xi'
\end{equation}
Here $\psi^{+},\psi$ are one particle field operators and $ h({\bf \xi},
{\bf \xi}')=h({\bf \xi}',{\bf \xi})$. This transformation gives canonical
transformation of the field operators $ \tilde{\psi}=\exp{(-i\hat{S})}\psi\exp{(i\hat{S})},
\quad\tilde{\psi}^{+}=\exp{(-i\hat{S})}\psi^{+}\exp{(i\hat{S})}$ which have
the form
\begin{equation}
\label{psi}
\tilde{\psi}({\bf r})=e^{i\hat{\alpha}({\bf r})}\psi,\quad 
\tilde{\psi}^{+}({\bf r})=
\psi^{+}({\bf r})e^{-i\hat{\alpha}({\bf r})}
\end{equation}
Here $\hat{\alpha}({\bf r})=2\int h({\bf r},{\bf \xi})\psi^{+}({\bf \xi},{\bf \xi})
d^2\xi$, electron spins assumed identical and spin indices are omited.
"Dressing" electrons with the "flux" is achieved by
\begin{equation}
\label{phase}
h({\bf(r-\xi)}=K\arctan\frac{r_y-\xi_y}{r_x-\xi_x}, \alpha({\bf r})=
2K\int\arctan
\frac{r_y-\xi_y}{r_x-\xi_x}d^2\xi
\end{equation}
It is assumed that operator $\hat{S}$ must be single valued for any realization
of electron coordinates (i.e. the electron density is the sum of $\delta$
functions) that gives integer $K$.
Transformed Hamiltonian is obtained by direct substitution of (\ref{psi}) into
the hamiltonian of interacting electrons:
$$ \tilde{H}= \frac{1}{2m}\int \tilde{\psi}^{+}(-i\hbar{\bf \nabla}-\frac{e}
{c}{\bf A}
+\frac{e}{c}{\bf \hat{a}})^2\tilde{\psi}d^2r+\frac{1}{2}\int V({\bf {r-r'}})\tilde{\psi}^{+}({\bf r})
\tilde{\psi}^{+}({\bf r'})\tilde{\psi}({\bf r'}\tilde{\psi}({\bf r})d^2r 
d^2r'$$
$$curl{\bf \hat{a}}=2K\Phi_{0}\tilde{\psi}^{+}({\bf r})\tilde{\psi}({\bf r})$$
where $\Phi_0$ is the flux quantum. This Hamiltonian is called Chern-Simons 
Hamiltonian with an 
artificial
6-fermion interaction due to the "dressing" with no small parameters. Further treatings of
this hamiltonian usually use mean field approximation for the operator of "effective" field
$\hat{a}$ :$curl{\bf \hat{a}}=2K\Phi_0 n_e$,where $n_e$ is average electron density. In this approximation
we have "effective" magnetic field additional to external one. At the integer fillings of Ll's in 
the total
field should be cyclotron gaps. That gives special  Jain's fractional fillings of the Ll's in 
the external magnetic
field $\nu=\frac{q}{1-2Kq}$ where $q$ is an integer. The choice $K=-1$ gives most part of observed
fractions. Performing any real calculations is very difficult task because mean field approximationis is used
quite
arbitrary and there is no small parameter to consider fluctuatios. Performed calculations of 
"Fermi liquid"
state at $\nu=1/2$ $(q\rightarrow\infty)$ give infinite effective mass [7].

However, the theory of FQHE can likely be developed on a different
physical basis that is associated with the existence of topological
textures stable to finite deformations. The topological
classification of multiparticle wave functions is a rather
complicated mathematical problem, and, to my knowledge, no simple
and, simultaneously, -effective definition of topological classes is
presently given. The classification of topological excitations is
well elaborated for a ferromagnetic 2D electron gas in a strong
magnetic field with filling $\nu = 1$ (skyrmions [8, 9]).

As an alternative to composite fermion approach it is possible to consider more simple canonical transformation with the same 
objectives but
conserving standard 4-fermionic interaction and connected with the topological textures of the vortex 
lattice type. I assume canonical transformation with the operator $\hat{S}$ of the form
\begin{equation}
\label{S'}
\hat{S} =\int \psi^{+}_{\mu}({\bf r})V_{\mu\nu}({\bf r})\psi_{\nu}({\bf r})d^2r
\end{equation}
I itroduce here some additional index for electron operators. It can be spin or some isospin. Really
2d electron system is obtained by filling of only one quantum state for the electron motion in 
transverse
direction. In that sense the electron states are ordered in transverse direction and isospin acquires 
only one value.
The energetical cost of other transverse states with different value of isospin 
define the size of an 
area with 
the rotated
isospin.Therefore we use some spinors $\psi_{\mu}$ and exact meaning of spinor index is not essential 
in the
further consideration. For simplicity I consider ordinary spin indices. The transformed spinors
$\chi_{\mu}=e^{-i\hat{S}} \psi_{\mu}
e^{i\hat{S}}$ have
the form $\chi_{\mu}({\bf r})= U_{\mu \nu}({\bf r})\psi({\bf r}), \chi^{+}_{\mu}({\bf r})=\psi^{+}_
({\bf r})U^{*}_{\mu \nu}({\bf r})$ of spinors after nonuniform rotation by the unitary matrix
$\hat{U}=\exp{i\hat{V}}= U_z(\alpha({\bf r})U_y(\beta({\bf r}) U_z(\alpha({\bf r})$ where
$\alpha,\beta,\gamma=\alpha$ are three Euler angles and the lower indices denote
the axis of 
the rotation.

 After the canonical transformation, the 
Lagrangian of interacting electrons takes the form (in the system of units 
where external magnetic field$B=1,l_B=1$,  and $\hbar =1$)
\begin{multline}
\label{L}
 L=\int \left[i\chi^{+}\frac{\partial \chi}{\partial t} -
\frac{1}{2m}\chi^{+}(-i\vec{\nabla+A_0+{\hat \Omega})^2\chi}\right]
d^2r +\\
 \frac{1}{2}\int
V(\vec{r}-\vec{r'})\chi^{+}(\vec{r})\chi^{+}(\vec{r'})
\chi(\vec{r'})\chi(\vec{r})d^2r d^2r'
\end{multline}

where 
$${\hat \Omega} = -i U^{+}{\bf \nabla}U={\bf \Omega}^l\sigma_l$$

$\sigma_l$ are Pauli matrices,
$${\bf \Omega}^z=\frac{1}{2}(1+cos\beta){\bf \nabla}\alpha$$
$${\bf \Omega}^x=\frac{1}{2}(\sin{\beta}\cos{\alpha}{\bf \nabla}\alpha-
\sin{\alpha}{\bf \nabla}\beta)$$
$${\bf \Omega}^y=\frac{1}{2}(\sin{\beta}\sin{\alpha}{\bf \nabla}\alpha+
\cos\alpha{\bf \nabla}\beta)$$
and $V({\bf r}-{\bf r}')$ is the Coulomb interaction. It is assumed
that $\gamma=\alpha$, because the angle $\gamma$ plays an
auxiliary role, eliminating the singularities of the matrix
$\hat{U}$. The spinors $\psi$ and $\psi^{+}$ are the electron-field operators
obeying the Fermi commutation rules. One can readily verify that
$\chi^{+}$ and $\chi$ satisfy the same commutation rules. The new
Lagrangian is formally equivalent to the initial one with ${\hat \Omega}=0$.
Hence, this Lagrangian gives electronic states corresponding to
${\hat \Omega}=0$, because one can always perform inverse transformation.
However, one may attempt to seek for any other states that are
characteristic of the Lagrangian with ${\hat \Omega}\neq0$]. 
This program can be
successfully implemented in the case where $U$ changes only slightly
at a distance of the order of magnetic length $l_B^2=\frac{\hbar c}{eB}$ 
 and all $l_B|{\bf \Omega}^l|$ are small. At large
distances, $\beta \rightarrow 0$, so the matrix $U$ only rotates spinors
around the $z$ axis, which aligns with the spin orientation in a
homogeneous ferromagnet and endows them a nontrivial phase. The
desired electronic state for operators $\chi$ and $\chi^{+}$ can be
obtained perturbatively for small ${\hat\Omega}$ from a uniform
ferromagnetic state for operators $\chi$. The existence of a
topological number
$$K = \frac{1}{2\pi}\int curl {\bf \Omega}^z d^2 r$$ 
which is determined by the number of revolutions through the
angle $\alpha(\vec{r})$ upon going around the infinite far contour, is a
nontrivial topological requirement. The value of $K\neq0$ 
precisely  defines the wave-function topological class and
makes the wave-function deformation into the trivial ferromagnetic
state with identical directions of all $\psi$ spinors impossible.
Thus, ${\hat \Omega}$ with different $K$ characterize the topologically
different classes of multiparticle wave functions. The condition
$\beta=\pi$ at the points of $\alpha(\vec{r})$ singularity (of the
polar-angle type) guarantees the absence of singularities for
${\hat \Omega}$. This approach was suggested in [10]; various physical
quantities were calculated in [11] in the leading order of
perturbation theory. The results for the syrmion energy coincide with those obtained by
other methods (see [7, 8]). The quantity $curl{\bf \Omega}^z$
plays the role of an additional effective magnetic field, and this field
is  the collective property of the multiparticle wave
function rather than the attribute of an individual electron. The
calculations of electron density, energy, and spin density can be, in
principle, carried out up to any order in the derivatives of matrix $U$.

This example demonstrates the method of determining isolated
topological excitations. However, this approach can be extended to
the analysis of the texture and a multiparticle wave function
corresponding to the finite density of topological number
$K$ on 2d plain. The analysis of arbitrary textures of this type for 
${\hat \Omega}$
involves great methodological difficulties and, likely, bears no
direct relation to the ground-state classification. We, therefore,
assume that these textures are near-periodic, so that the average-spin
field is periodic. Essentially I want to construct periodic vortex lattice.
 Let us consider an elementary cell. We assume that the
average spin vector at the elementary cell boundary has a constant value and
is aligned with the $z$ axis in the spin space. Thus, the angle
$\beta$ is assumed to be a periodic function in 2d plane, with $\beta =0$ 
at the elementary cell boundaries. The angle $\alpha$ is assumed to
possess vortex singularity at some point inside each unit cell, for
which we assume that $\beta = \pi$ in order to eliminate the singularities 
of ${\hat \Omega}({\bf r})$. One can set, for example, 
$\alpha=\sum\alpha_i$,
where the summation goes over
all elementary cells and $\alpha_i({\bf r})$ is the polar angle centered inside
the $i$ elementary cell.The detailed form of $\alpha({\bf r})$ is not very important because
$\hat{\Omega}$ form some vector potential and gradient transformation
can eliminate part of this matrix. The circulations of ${\bf \Omega}^k$ along the
sides of elementary cell are easily calculated and give 
$\oint {\bf \Omega}^zd{\bf l}=2\pi K$,
$\oint{\bf\Omega}^{x,y}d{\bf l}=0$,where $K$ is the winding number for $\alpha$ inside each 
cell. Thus we have constant total flux through each elementary cell for positive spin in direction 
of external magnetic field $\Phi=Ba_c+\Phi_0 K$ ($a_c$ is elementary cell area) and 
periodical "effective" magnetic field due to 
$curl{\bf \Omega}^{x,y}$ with zero total flux. The interaction in Lagrangian 
(\ref{L}) is translation
invariant. 

 Therefore, each cell is characterized by the same
topological number
$K =\frac{1}{2\pi}\int curl {\bf \Omega}^z d^2r$
that specifies the integer number of the flux quanta for the additional
effective magnetic field with the average value $B_{eff}=\frac{K}{a_c}$
over the sample area. Taking
ferromagnetic $\chi$ and $\chi^{+}$ as an approximation, the
average spin ${\bf n}({\bf r})$ gives the $K$-fold mapping of any elementary
cell onto unit sphere.  Although the sum $\alpha=\sum\alpha_i$
over all cells is, formally, a
periodic function, it diverges. Since only $\sin\alpha$ and
$\cos\alpha$ enter the expression for ${\hat \Omega}$, the modulo $2\pi$
convergence is sufficient. I will adopt, without proof, that 
 ${\bf \Omega}^{x,y}$ can be regularized 
in a periodic manner.

It is not my intention to calculate electron energy in such textures.
This is a rather complicated problem for elementary cell sizes of the order
of magnetic length, for which the gradient expansion in ${\hat \Omega}$ is
impossible. My goal is to classify the electronic states with the aim
of determining certain special density values that correspond to the
ground states separated by a gap from the excited states. The problem
of numerical calculation of the gap can be posed after the
classification of ground states.

We have, in fact, a system of interacting electrons in a periodic
effective magnetic field (the sum of the external magnetic field and
a periodic vortex "magnetic field" in elementary
 cells) with nonzero average.
The corresponding transformation group consists of the magnetic
translations and is the projective representation of the conventional
translation group. According to the well-known analysis (Brown, Zak;
see, e.g., [13]) for noninteracting electrons, the band spectrum is
regular only for a rational number of flux quanta. The irrational
number of quanta or a rational number with large coprime
numerator and denominator gives a highly irregular structure
with the allowed and forbidden bands dense in a certain energy
region. One can assume, in the spirit of the Fermi-liquid theory,
that the interaction does not affect these spectral features.
Restricting oneself to the rational fluxes we get

\begin{equation}
\label{flux}
Ba_c+K\Phi_0=\frac{p}{q}\Phi_0
\end{equation}
where $p,q$ are integers and
 one obtains $a_c =\frac{p-Kq}{q}\frac{\Phi_0}{B}$ 
for the elementary
cell area.
The total number of states per unit area, with one electron per elementary
cell, determines the electron density $n_e=\frac{B}{\Phi_0}\frac{q}{p-qK}$
and must correspond to the filled set of bands
obtained from $\frac{S}{a_c}$ states in the absence of the average magnetic field,
though in a periodic potential or periodic magnetic field with the zero average 
with the period specified by the elementary
cell. Here, $S$ is the sample area. Simple analysis states [13] that
this initial band is split into $q$ subbands, each being (odd $q$) $q$-
fold or (even $q$) $q/2$-fold degenerate, and with the fraction of the
number of states in each subband being (odd $q$)equal to $1/q^2$ or (even $q$)
to $2/q^2$. However, the total number of states in all subbands is
$S/a_c$. One can assume that, even in the presence of
interaction, these states are separated from the higher-energy states
by the greatest gap. The structure of inner forbidden bands is
irrelevant because all lower-lying states are filled. Note that the
evenness of the $K$ number is immaterial, because
the Fermi commutation rules
for the operators $\chi$ and $\chi^{+}$ are fulfilled automatically and
have no relation to the topological number $K$. The occurrence of any
specific numbers of vortex-field flux quanta can be dictated  
by the ground-state energy . The observed fractions in FQHE correspond to the following tables

\vspace{5mm}
 
$K=-2,\;\;p=1$

\vspace{3mm}

\begin{tabular}{|c|c|c|c|c|c|c|c|c|c|}
\hline
$q$ & 1&2&3&-5&-2&-3&-4&4&$\infty$\\
\hline 
$\nu$
&$\frac{\mathstrut 1}{\mathstrut 3}$
&$\frac{2}{5}$
&$\frac{3}{7}$
&$\frac{5}{9}$
&$\frac{2}{3}$
&$\frac{3}{5}$
&$\frac{4}{7}$
&$\frac{4}{9}$
&$\frac{1}{2}$\\
\hline
\end{tabular}

\vspace{5mm}

That fractions correspond to celebrated Jain's rule. Half filling of the 
Landau level $n_e=\frac{B}{2\phi_0}$ in the external field corresponds to
vanishingly small effective magnetic field (zero number of flux
quanta per elementary cell).

Other observed fractions correspond to
\vspace{5mm}

$K=-1,\;\;p=1$
\vspace{3mm}

\begin{tabular}{|c|c|c|c|}
\hline
$q$&-4&4&2\\
\hline
$\nu$&$\frac{\mathstrut 4}{\mathstrut 3}$&$\frac{4}{5}$&$\frac{2}{3}$\\
\hline
\end{tabular}

\vspace{5mm}
where one have double of the fraction 2/3, and

\vspace{5mm}
$K=-1,\;\;p=2$

\vspace{3mm}
\begin{tabular}{|c|c|c|c|c|}
\hline
$q$&-7&-5&5&2\\
\hline
$\nu$&$\frac{\mathstrut 7}{\mathstrut 5}$&$\frac{5}{3}$&$\frac{5}{7}$&$\frac{1}{2}$\\
\hline
\end{tabular}

\vspace{5mm}
here one have not observed double of the fraction 1/2 with the gap
($B_{eff}\neq0$).

Thus, I have reproduced the key statement of the theory of composite
fermions (Jain's rule) and obtained the explanation of all observed fractions
at moderate Lls filling in  unified frame without any hierarchial schemes.
 Of course, these results are quite crude and, in some points
hypothetical. The energy gap, the properties of elementary
charge excitations, and the conductivity calculations, as well as the
analysis of different $K$ and $p,q$ values are still open questions, and
the approach to these problems is as yet unclear.The preliminary results where
published in [14].

I am grateful to V.G. Dolgopolov, V.F. Gantmakher, V.B. Timofeev, and
M.V. Feigelman for helpful discussions. This work was supported by
the Russian Foundation for Basic Research, the Program for Supporting 
Scientific Schools, and the State Contract
40.020.1.1.1165 of the Ministry of Science of the Russian Federation.

REFERENCES

[1] The Quantum Hall Effect, Ed. by R. Prange and S. M. Girvin
(Springer, New York, 1987; Mir, Moscow, 1989).

[2] New Perspectives in Quantum Hall Effects, Ed. by S. Das Sarma
and A. Pinczuk (Wiley,  1997).

[3] R. B. Laughlin, Phys. Rev. B {\bf22}, 5632 (1981).

[4] R. B. Laughlin, Phys. Rev. Lett. {\bf50}, 1395 (1983).

[5] J. K. Jain, Phys. Rev. Lett. {\bf63}, 199 (1989).

[6] J. K. Jain, Phys. Rev. B {\bf41}, 7653 (1990).

[7] B. I. Halperin, P. A. Lee, and N. Read, Phys. Rev. B {\bf47}, 7312
(1993).

[8] S. Sondhi, A. Kahlrede, S. Kivelson, and E. Rezayi, Phys. Rev. B
{\bf47}, 16 418 (1993).

[9] K. Moon, N. Mori, Kun Yung, S.M.Girvin, A.H.Macdonald, L.Zheng,
D.Ioshioka, S.Zhang,  Phys. Rev. B {\bf51}, 5138 (1995).

[10] S. V. Iordanski, S. G. Plyasunov, and I. V. Fal'ko, Zh. Eksp.
Teor. Fiz. {\bf115}, 716 (1999) [JETP {\bf88}, 392 (1999)].

[11] S.Brener,S. V. Iordanski and A. B. Kashuba,  Phys.
Rev. B {\bf 67}, 125309(2003) 

[13] E. M. Lifshitz and L. P. Pitaevski, Course of Theoretical
Physics, Vol. [IX]: Statistical Physics, Part 2(Nauka, Moscow, 1978;
Pergamon, New York, 1980)

[14] S.V.Iordanski,Pisma ZhETF v{\bf77} iss.5, p292 (2003)

\end{document}